\documentclass[a4paper]{EslabStyle}
\usepackage{epsfig}

\newcommand{\beq}{\begin{equation}}
\newcommand{\eeq}{\end{equation}}
\def\etal{{\it et al.}}
\def\mpc {h^{-1} {\rm Mpc}}
\begin{document}
\title{Simulated Submillimetre Galaxy Surveys}

\author{David H. Hughes
\& Enrique Gazta\~{n}aga}
\institute{Instituto Nacional de Astrofisica, Optica y Electronica (INAOE),
Tonantzintla, Apdo Postal 216 y 51, 7200, Puebla, Mexico}

\maketitle

\begin{abstract}
Current submillimetre surveys are hindered in their ability to reveal
detailed information on the epoch of galaxy formation and the
evolutionary history of a high-redshift starburst galaxy
population. The difficulties are due to the small primary apertures
($\rm{D < 15}$-m) of existing submillimetre telescopes and the limited
sensitivities of their first generation of bolometer cameras.  This
situation is changing rapidly due to a variety of powerful new
ground-based, airborne and satellite FIR to millimetre wavelength
facilities. Improving our understanding of the luminosity and
clustering evolution provides the motivation for conducting
cosmological submillimetre and millimetre surveys.  It is therefore
important that we quantify the limitations of the future surveys and
the significance of the results that can be drawn from them.  In this
paper we present simulated surveys which are made as realistic as
possible in order to address some key issues confronting existing and
forthcoming surveys.  We discuss the
results from simulations with a range of wavelengths (200$\mu$m
-- 1.1\,mm), spatial resolutions (6\arcsec -- 27\arcsec) and flux
densities (0.01--310 mJy).  We address how the measured source-counts
could be affected by resolution and confusion, by the survey
sensitivity and noise, and by the sampling variance due to clustering
and shot-noise.
\end{abstract}

\section{Introduction}

By early 2001 the initial extensive programme of extragalactic SCUBA
(850$\mu$m) surveys conducted on the 15-m JCMT will be completed.  The
targets of these sub-mm surveys include lensing clusters (Smail \etal\
1997), the Hubble Deep Field (Hughes \etal\ 1998), the Hawaii Deep
Survey fields (Barger \etal\ 1998, 1999), the Canada-France Redshift
Survey fields (Eales \etal\ 1999, Lilly \etal\ 1999) and the UK 8\,mJy
SCUBA surveys of the ISOPHOT ELAIS and Lockman Hole field (in prep.)  These
blank-field sub-mm surveys, which cover areas of $\sim 0.002 -
0.2$\,deg$^{2}$, allow a measure of the $850\mu$m source-counts
between flux densities of 1--15mJy.  The relatively small survey areas
are due to the low mapping speed (a combination of the field-of-view
and sensitivity of the bolometer array, and the necessary overheads to
fully sample the focal-plane with $\rm >0.5F\lambda$ feedhorns, {\it
e.g.} 16 separate secondary-mirror positions are required to fully
sample the SCUBA array). The restricted 850$\mu$m flux range in the
measured source-counts is the result of a source-confusion limit of
$\sim 2$\,mJy (due to a small primary aperture or map resolution) 
at faint flux densities and the small survey area at the bright end,
given the intrinsic low source-density of bright sub-mm
sources ({\it e.g.}  $N[S_{850_{\mu m}} > 20\rm \,mJy] < 70~~
sources/deg^{2}$).
  


The limitations on the accuracy of our understanding of high-z galaxy
evolution from sub-mm surveys has been described elsewhere (Hughes
2000). To improve the constraints on the competing evolutionary
models, future sub-mm and mm surveys must extend their wavelength
coverage, increase their survey area and sensitivity. To quantify the
advantages of these forthcoming surveys we have simulated the
extra-galactic sky at various sub-mm and mm wavelengths
(200-3000$\mu$m), spatial resolutions ($\theta_{FWHM} \sim
1-120$\arcsec) and sensitivities ($S >0.01$ mJy). 
We include in the simulations
contributions from CMB primary fluctuations, S-Z clusters, an evolving
extra-galactic population of star-burst galaxies and high-latitude
galactic-plane cirrus emission.  These simulated surveys allow us to
compare the measured source-counts for a wide variety of current and
future FIR and sub-mm/mm telescopes ({\it e.g.} SIRTF, FIRST, MAP, 
PLANCK, BLAST,
SOFIA, CSO, JCMT, LMT, GBT, ALMA).

We discuss these multi-wavelength simulations
in more detail elsewhere (Hughes \& Gazta\~{n}aga 2000).  
In this paper we concentrate on illustrating the
effects of galaxy clustering and low spatial resolution on the
measured source-counts at 1100$\mu$m, 850$\mu$m, 350$\mu$m and 200$\mu$m. 

\section{Simulations}

The starting point for our simulations is a $P^3M$ N-body simulation
with $200^3$ particles in a $600 \mpc$ box that has the same matter
power spectrum as that measured for APM Survey galaxies (see
Gazta\~{n}aga \& Baugh 1998, and references therein).  Thus, the
2-point correlation in the (dark) matter particles is identical to the
one measured in the local ($\bar{z} \simeq 0.15$) galaxy universe.
Using a single N-body output (as opposed to a light-cone output)
corresponds to having the matter clustering pattern fixed in comoving
coordinates (stable clustering). If redshift evolution of the two-point
correlation function $\xi_2(r,z)$ is parametrized as:

\beq
\xi_2(r,z)= (1+z)^{-(\epsilon+3)} ~\xi_2(r), 
\label{evxi2}
\eeq
then, for stable clustering we have $\epsilon \simeq 0$ at small
scales ({\it e.g.} see Gazta\~naga 1995).
This has less clustering evolution than pure matter
gravitational clustering ({\it e.g.} linear or non-linear growth,
where $\epsilon \simeq 1$) and
might describe some models in which galaxies are identified with
high-density matter peaks (peaks move less than particles, which
results in less evolution).  This might be adequate for star-forming
galaxies if we are selecting rarer (more massive) objects as we go to
higher redshifts.  It is in agreement with the strong clustering
observed in Lyman-break galaxies at $z \simeq 3$, which is comparable
to the clustering of present-day galaxies (see Giavalisco \etal\,
1998).

We next implement a model for the evolution of the galaxy population
and set the geometry of a light-cone observation.  To relate spatial
coordinates $r$ and observed redshifts $z$ in the simulation output we
use a spatially flat metric with the critical density $\Omega=1$ and
zero cosmological constant (other cosmological parameters will be
considered elsewhere).  We transform the N-body matter simulation into
a mock galaxy catalogue of angular and redshift positions by the
following steps: 
\begin{itemize}
\item i) select an arbitrary point in the simulated box to
be the local `observer'; 
\item ii) apply a given survey angular mask ({\it
e.g.} a square with 1 degree on a side); 
\item iii) include in the mock
catalogue a simulated particle at redshift $z$ from the observer with
probability given by some selection function $\psi(z)$; 
\item iv) assign a luminosity $L=L(\lambda)$ to this particle
according to the luminosity
function $\phi[L,z]$ for each one of the observer filters $\lambda$,
at the corresponding particle redshift $z$. 
\end{itemize}

We replicate the N-body
(periodic) simulation box to cover the total extent of the survey (up
to $z=6$, beyond which we impose an exponential cut-off in the expected
number of galaxies).  By comparing the results from different box
sizes we have verified that this replication of the box does not
introduce any spurious correlations on large scales.  The selection
function $\psi(z)$ is the normalized probability that a galaxy at
redshift $z$ is included in the catalogue, and is proportional to the
estimated number of galaxies at this coordinate:

\begin{equation}
\psi(z) \propto \int_{L_1(z)}^{L_2(z)} ~dL~\phi[L,z] \label{psi}.
\label{sel}
\end{equation}
where $\phi[L,z]$ is the luminosity function and $L_1(z)$ and $L_2(z)$
are the luminosities corresponding to the lower and upper limits in
the range of apparent fluxes used to build the galaxy sample or
catalog under study.  Thus our prescription for galaxy formation is
that, on average, the probability to find a galaxy is simply
proportional to the (dark) matter densities, so that the number
density of galaxies automatically determines how massive are the
matter peaks that our galaxies are tracing. Other ``biasing'' schemes
can also be implemented, but note that different models for galaxy
formation can give different clustering evolutions for a given
$\phi[L,z]$, so that both of these evolutionary patterns need to be
better constrained by the future data if we want more realistic
simulations.

\begin{figure}
\centerline{\epsfxsize=8truecm 
\epsfbox{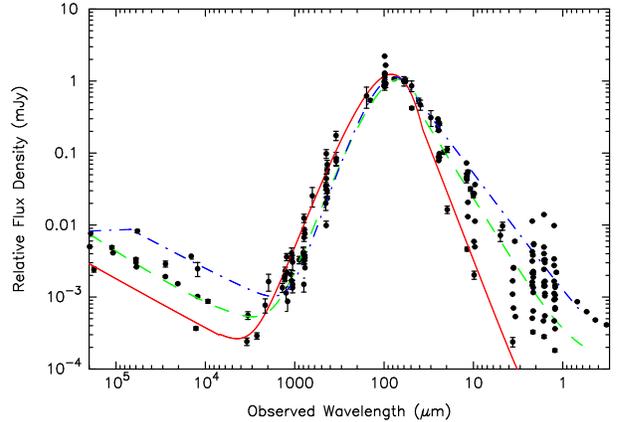}}
\caption{Radio to IR SEDs of low-z starbursts, quasars, ULIRGS,
Seyferts normalised at 60$\mu$m. The curves represent fits to the SEDs
of Arp220 (solid), M82 (dashed) and Mkn231 (dashed-dotted).  Since the
rest-frame SEDs show a large dispersion in the spectral slope at
$\lambda < 60\mu$m then only the number-counts of the highest-z
sources ($z > 3$) in the shortest wavelength simulations
(200-300$\mu$m) are sensitive to the exact choice of galaxy SED type
or its redshift evolution.  }
\label{SED}
\end{figure}

To include the star-formation history of the galaxies in our
simulations we have selected models describing the SEDs 
shown in  Figure \ref{SED} to extrapolate the local IRAS
$60\mu$m local luminosity function $\phi[L,z=0]=\phi[L/L^*]$ 
(Saunders \etal\ 1990) to longer wavelengths. 
We have introduced an exponential cutoff for objects fainter than
$L= 10^8 L_{\odot}$ and adopted a model of pure luminosity
evolution, $\phi[L,z]=\phi[L/L^*]$, with the following redshift
dependence: $L^*(z) = (1+z)^{3}~L^*(z=0)$ for $0 < z < 2.2$; $L^*(z) =
L^*(z=2.2)$ for $2.2 \le z < 6$; an exponential cutoff at $z >
6$. 

In this paper we use a single SED model of Arp220 for all galaxies in
our sample.  As will be shown below (Figure \ref{selwall}), using a
different SED makes no significant difference in the number counts or
redshift distributions even for bright sources at the shortest
submillimetre wavelengths. We have also assumed no evolution for the
SED: if the evolution occurs within the spread in the models shown in
Figure \ref{SED}, this has little effect for $\lambda \simeq
200-1100\mu$m.

\begin{figure*}[t]
\centerline{\epsfig{file=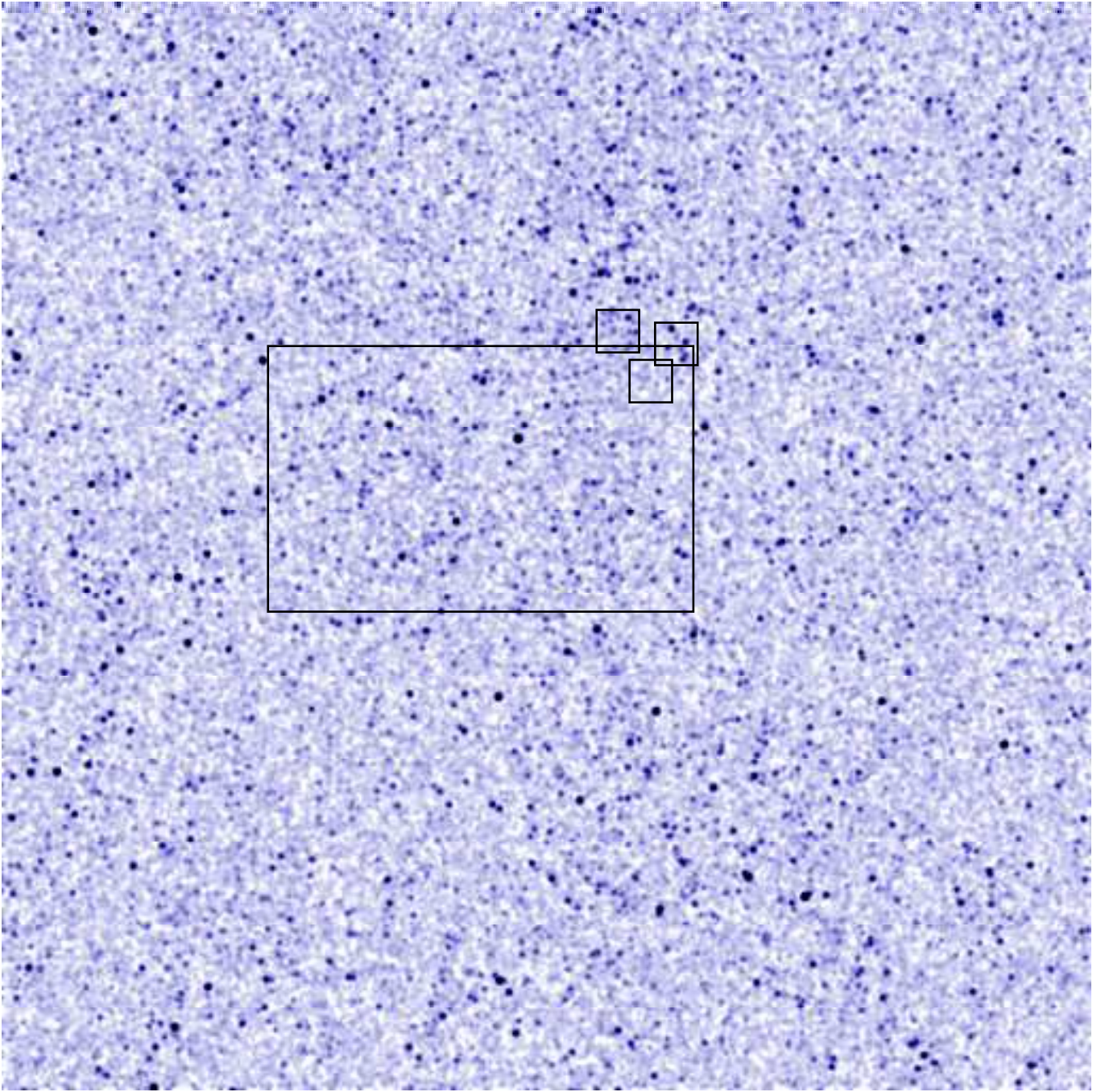, width=18cm}}
\caption{A confusion-limited 
($1\sigma = 0.5$\,mJy) JCMT 1\,deg$^{2}$ area simulated survey at 
850$\mu$m  with $\theta_{FWHM} = 15$\arcsec. The large and small boxes
show the 0.1\,deg$^{2}$ area and the  HDF-like regions shown in Figures \ref{maps}
\& \ref{hdf} respectively.}
\label{large850}
\end{figure*}

\begin{table*}
\begin{center}
\begin{tabular}{cc|cc|cc}\scriptsize
& LMT Surveys & \multicolumn{2}{|c|}{NEFD = 10 mJy sec$^{1/2}$ , $\eta \sim 100\mu$m} &
\multicolumn{2}{|c}{NEFD =  4 mJy sec$^{1/2}$ , $\eta \sim 70\mu$m} \\ \hline
Survey & Area (sq. arcmins) & 3-$\sigma$ depth & N\,($5-\sigma$ galaxies) & 
3-$\sigma$ depth & N\,(5-$\sigma$ galaxies) \\ \hline
A & 6    & 0.1 mJy    &  18    &   0.05 mJy  &  33   \\
B & 350  & 0.9 mJy    &  204   &   0.4  mJy  &  515  \\
C & 1000 & 1.6 mJy    &  320   &   0.6  mJy  &  1100 \\
D & 3600 & 3.0 mJy    &  400   &   1.2  mJy  &  1600 \\ 
\end{tabular}
\end{center}
\caption{Number of 5-$\sigma$ galaxies detected in alternative 50-hour
BOLOCAM 1.1~mm surveys with  a conservative Noise
Equivalent Flux Density (NEFD) of NEFD = 10 mJy sec$^{1/2}$ at 1.1~mm, 
during the initial commissioning phase ({\it e.g.} with a
primary aperture r.m.s. surface accuracy of $\eta \simeq
100 \mu$m r.m.s.) of the LMT; and later during routine
operation with an improved surface accuracy (of $\eta \simeq 70 \mu$m) and
1.1\,mm NEFD =  4 mJy sec$^{1/2}$.
BOLOCAM sensitivities allow for overheads. Surveys A and B
represent LMT surveys similar in area to the SCUBA survey of the HDF and
the UK 8~mJy wide-area survey respectively. Surveys C and D represent
future wide-area LMT surveys.}
\label{table:sur}
\end{table*}

\begin{figure*}
\centerline{\epsfig{file=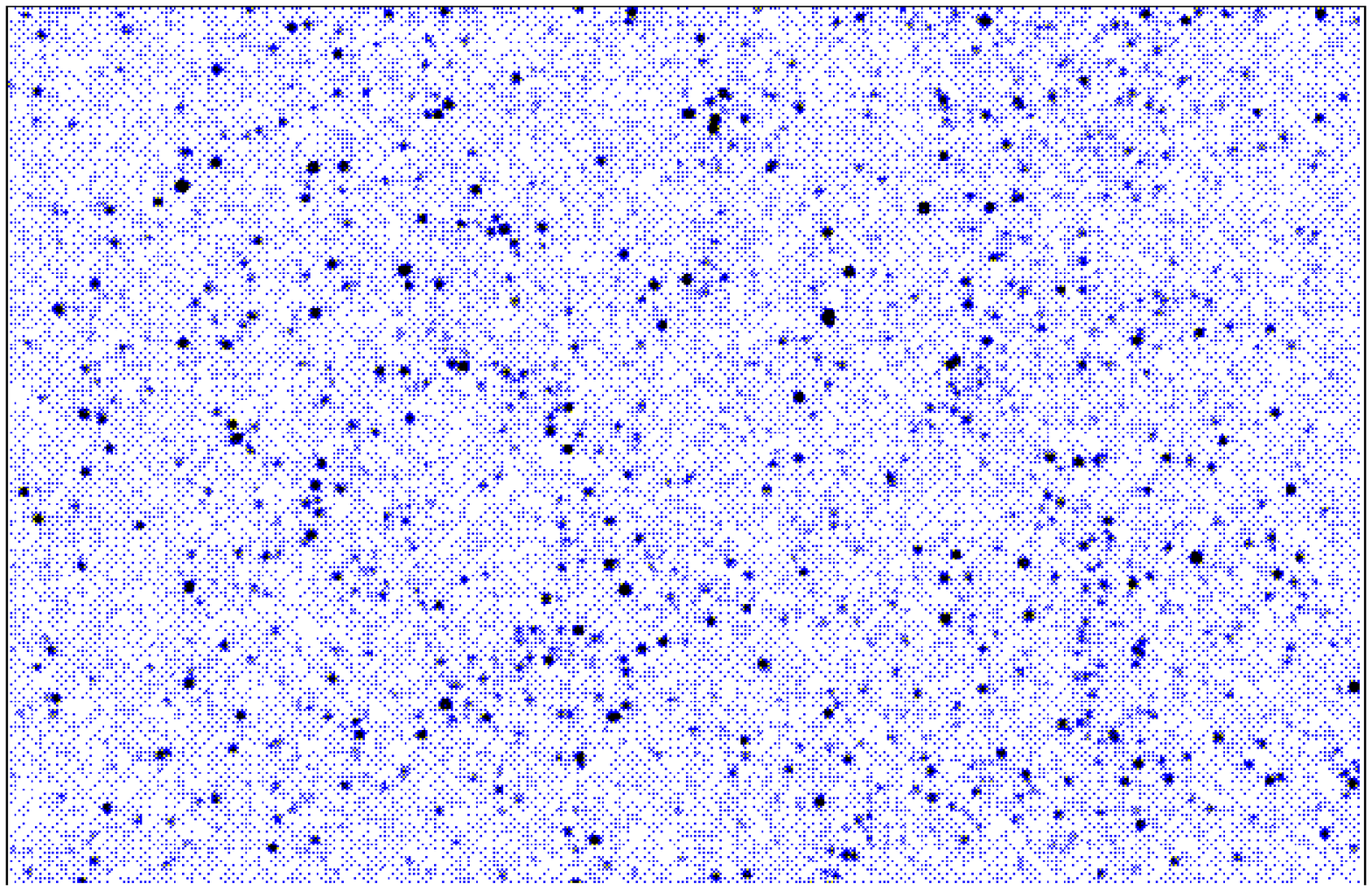, width=8cm}~
\epsfig{file=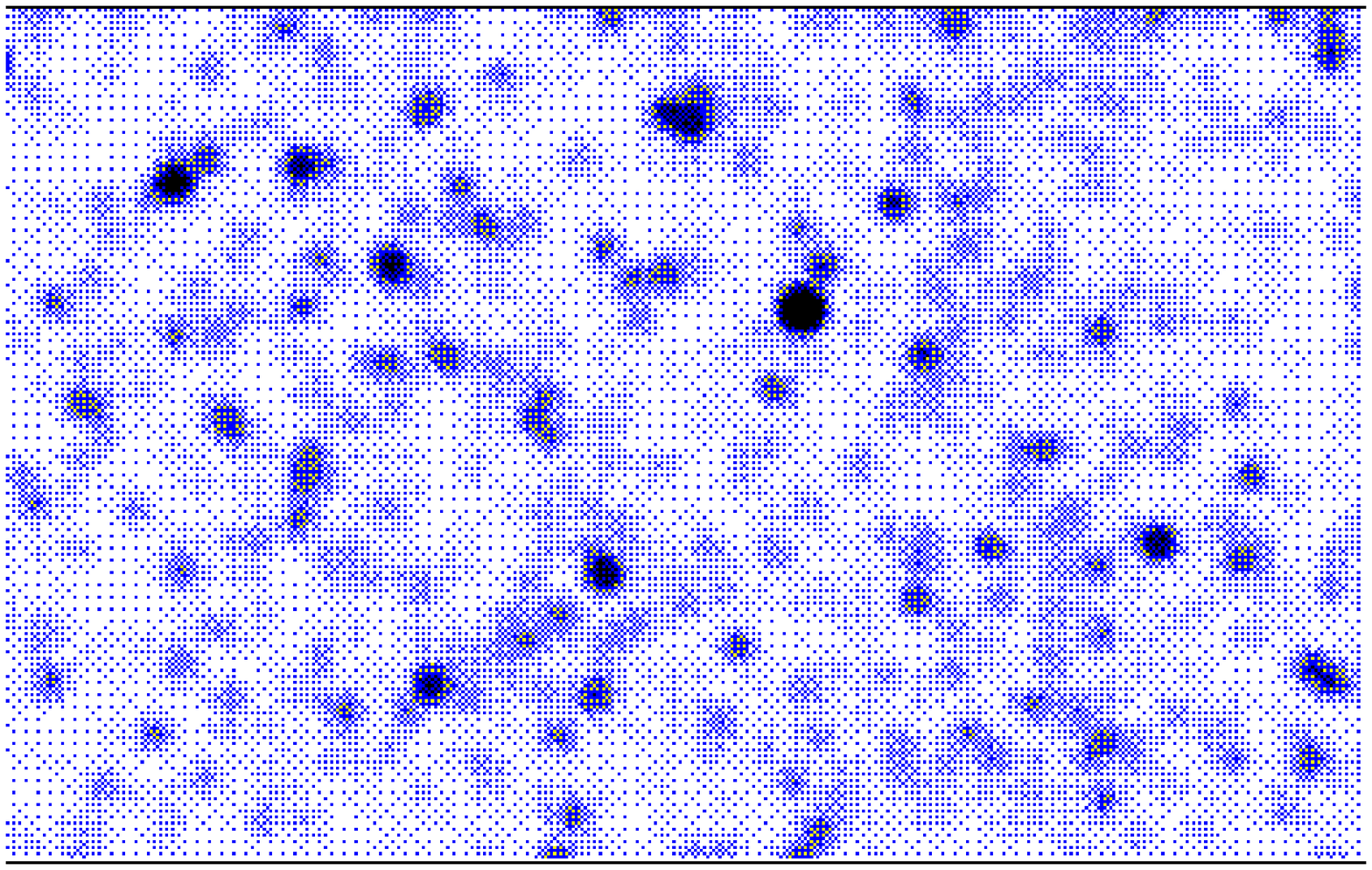, width=8cm}}
\centerline{\epsfig{file=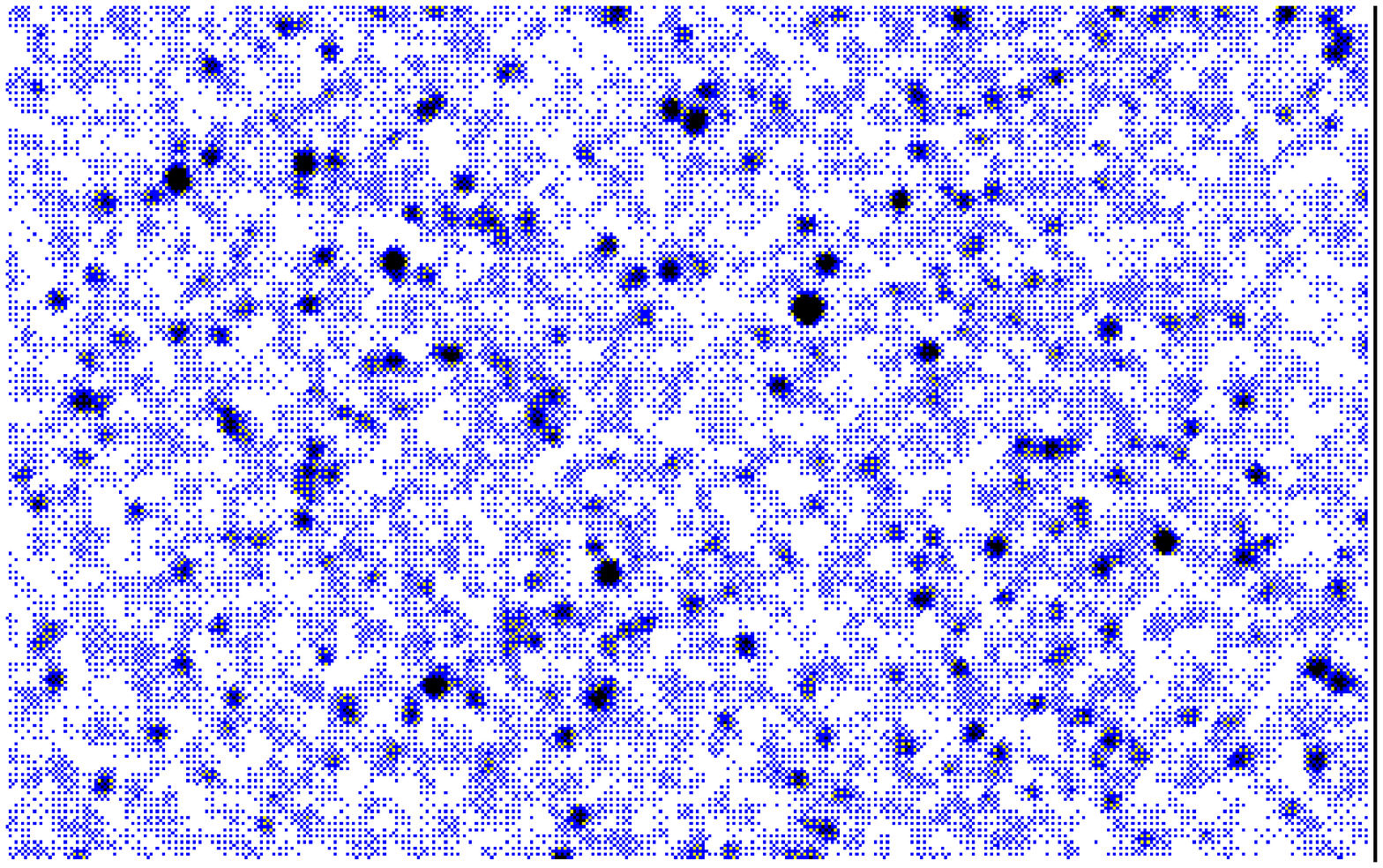, width=8cm}~
\epsfig{file=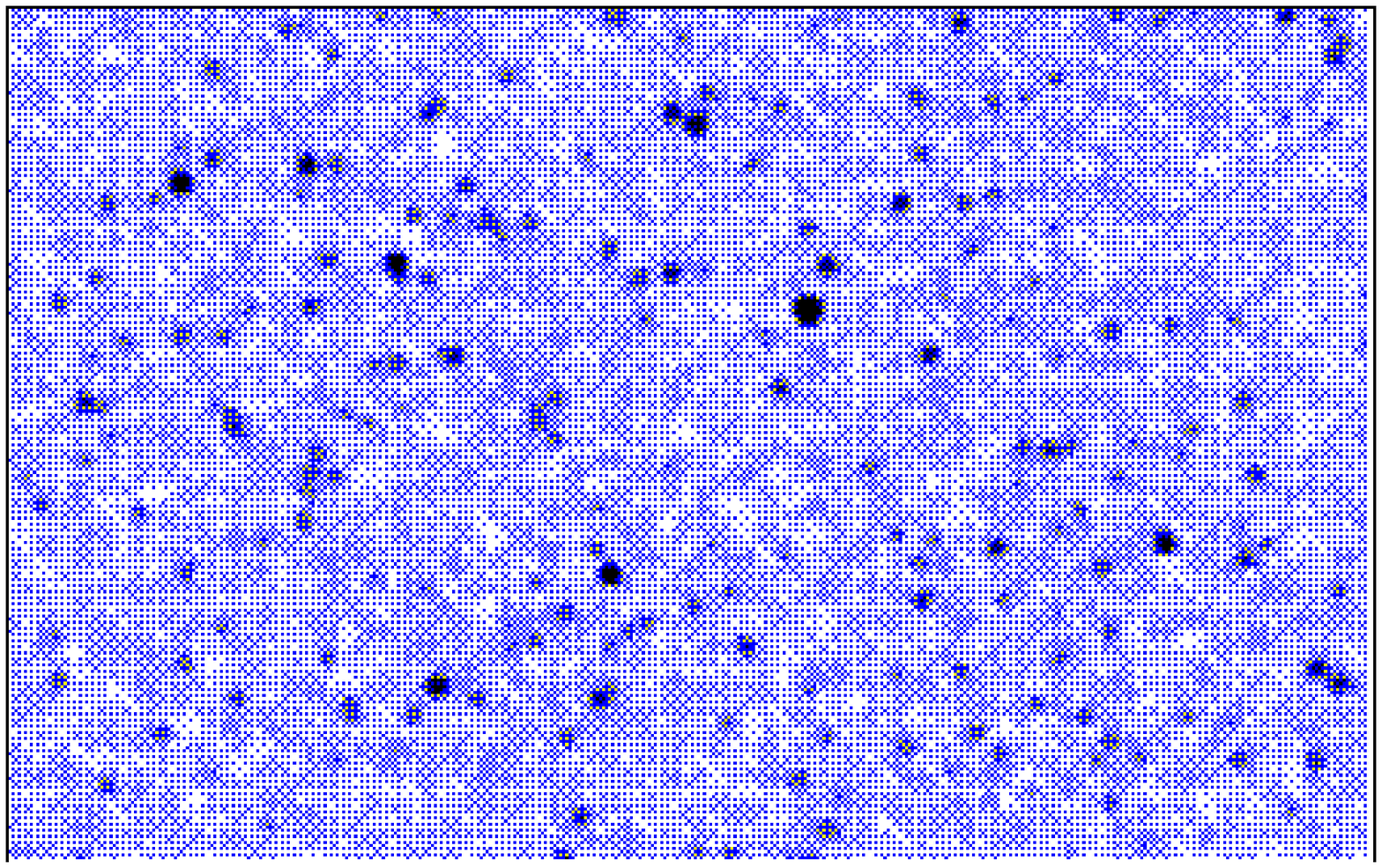, width=8cm}}
\centerline{\epsfig{file=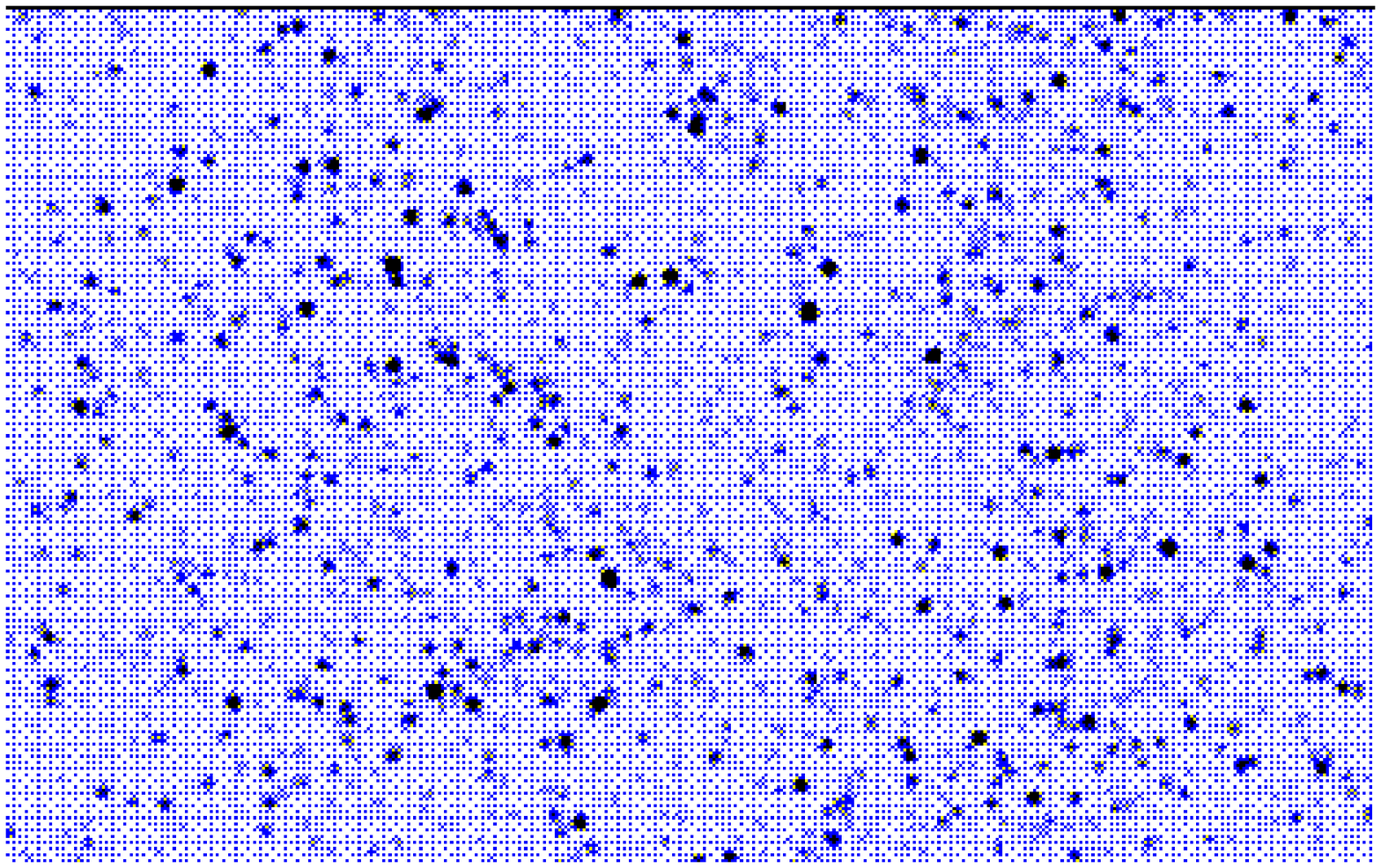, width=8cm}~
\epsfig{file=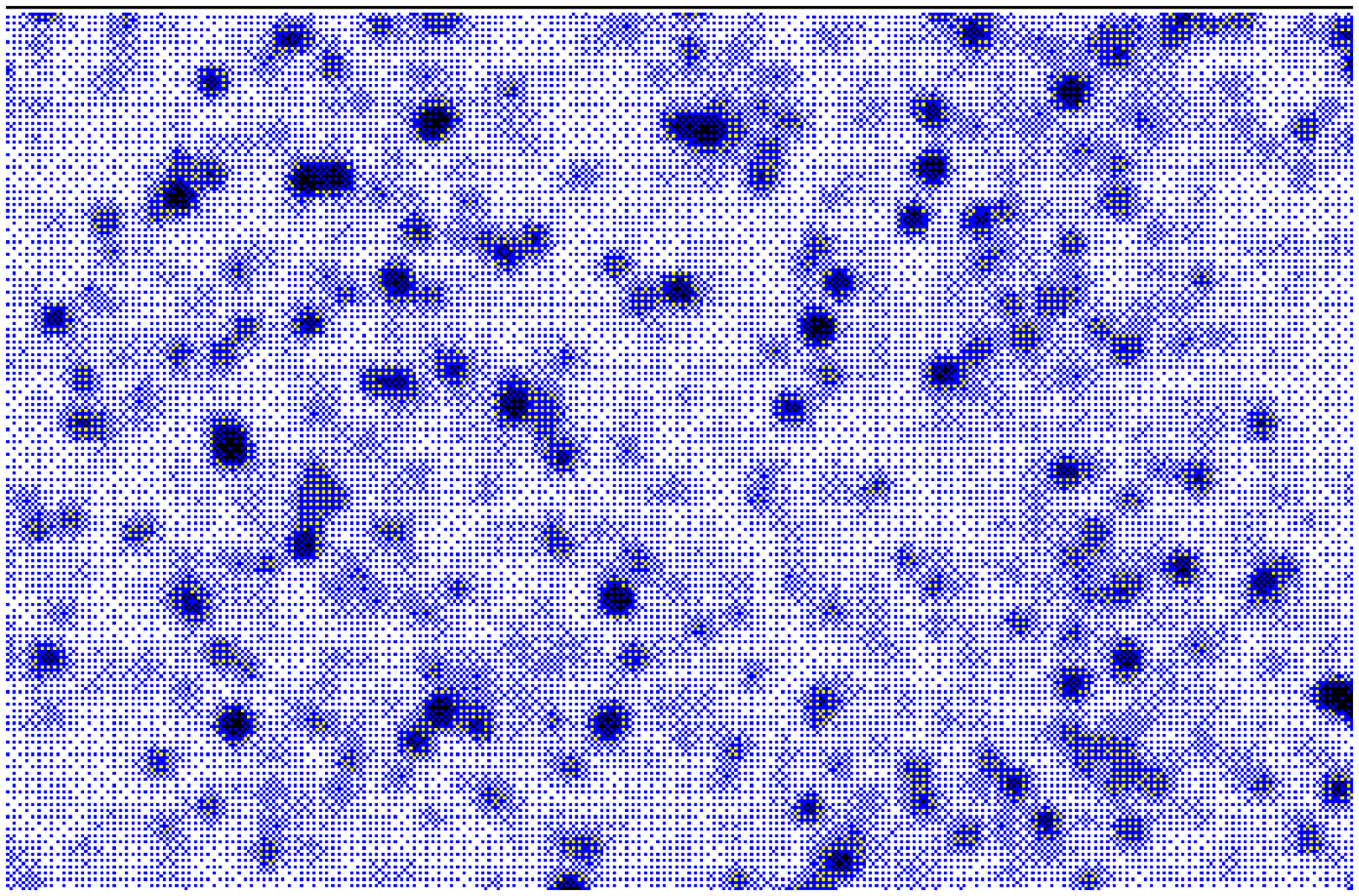, width=8cm}}
\caption{Simulated surveys covering 0.1\,deg$^{2}$: A
confusion-limited ($3\sigma = 0.05$\,mJy) LMT survey at 1100$\mu$m
with $\theta_{FWHM} = 6$\arcsec (top-left); a $\theta_{FWHM} =
27$\arcsec low resolution 1100$\mu$m survey (top-right); a
$\theta_{FWHM} = 15$\arcsec deep ($1\sigma = 0.5$\,mJy) 850$\mu$m
survey (middle-left); a $\theta_{FWHM} = 15$\arcsec shallow ($1\sigma =
2.5$\,mJy) 850$\mu$m survey (middle-right); a confusion-limited
($\theta_{FWHM} = 9$\arcsec) 350$\mu$m survey (bottom-left); a
confusion-limited ($\theta_{FWHM} = 25$\arcsec) 200$\mu$m survey
(bottom-right). These figures look much better in
http://www.inaoep.mx/~gazta/simsubmm.ps}
\label{maps}
\end{figure*}

Thus, after these transformations we end up with a galaxy catalogue of
angular positions, redshifts and luminosities, for each of the
observer filters $\lambda$. We next smooth the resulting projected
distribution with a spatial resolution appropriate to the
observational configuration we are simulating: $\theta_{FWHM} \simeq 1.2~
\lambda/\rm D$ where $\rm D$ is the telescope aperture. Finally we add
white noise ${\cal N}$ to the angular catalogue to emulate the
signal-to-noise ${\cal S/N}$ appropriate for the integration time,
$t$, the instrumental sensitivity and observational conditions (Noise
Equivalent Flux Density, NEFD) of the mock survey we want to
emulate.

We describe here the results from simulated surveys at 1.1\,mm,
850$\mu$m, 350$\mu$m and 200$\mu$m with spatial resolutions of
6\arcsec, 15\arcsec, 9\arcsec and 25\arcsec corresponding to surveys
on the 50-m LMT (http://binizaa.inaoep.mx), 15-m JCMT, 10-m CSO and
2-m BLAST (http://www.hep.upenn.edu/blast/) respectively.  The full
simulations have a flux dynamic range of $S_\lambda = 0.01 - 310$\,mJy
at all wavelengths and cover an area of 1 sq. degree.  Subsets have
been extracted to determine the source-counts from surveys comparable
in area (6--400\,arcmin$^{2}$, Table \ref{table:sur} - survey A and B)
to the SCUBA surveys of the Hubble Deep Field and Hawaii Deep Fields,
the lensing cluster survey, Canada-France Redshift survey fields and
the UK 8\,mJy ELAIS and Lockman Hole survey.  Additionally larger-area
surveys (0.3--1.0\,deg$^{2}$, Table 1 - surveys C and D) more
appropriate to the LMT and BLAST are also considered.

Figure \ref{large850} shows a large projected (1\,deg$^{2}$) map at
850$\mu$m with some identified sub-regions within which  we perform the
source-counts analysis. Although there is significant clustering 
in this figure, the overall distribution is already quite homogeneous. 
Figure \ref{maps} shows a comparison of maps at different
resolutions and wavelengths from the 0.1\,deg$^{2}$
region in the center of the larger map.

\begin{figure*}[ht]
\centerline{\epsfig{file=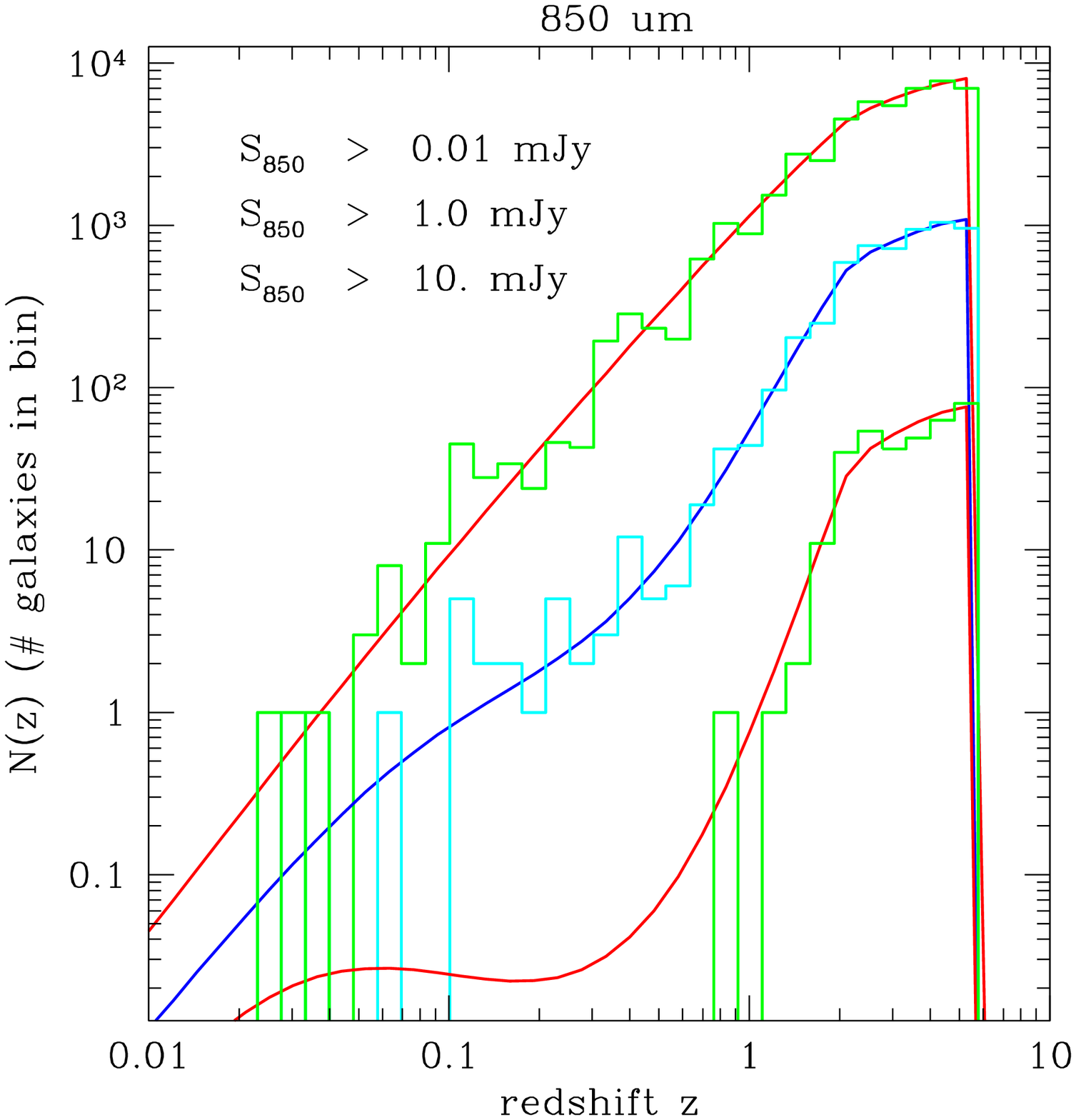, width=8cm}
\epsfig{file=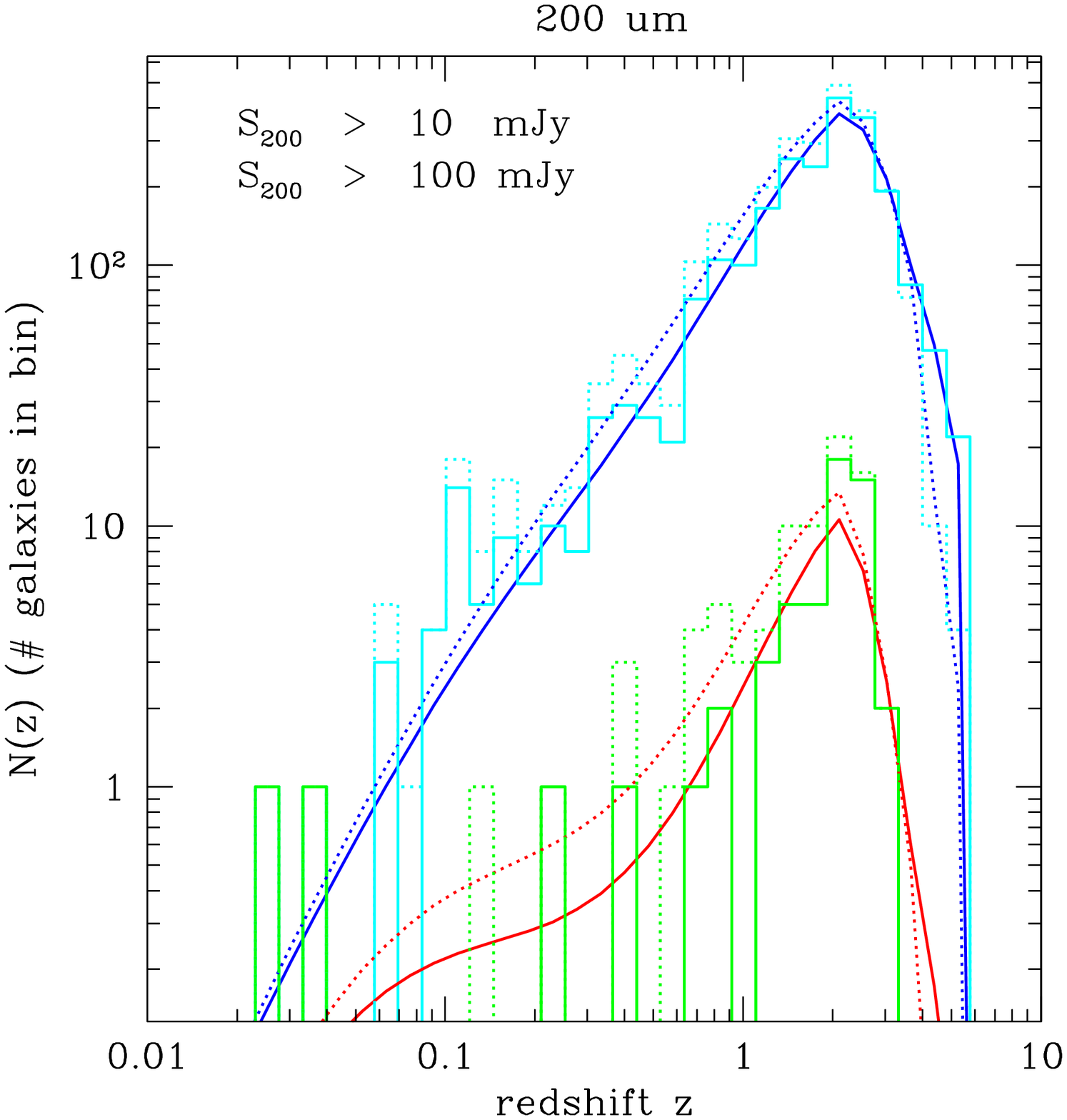, width=8cm}}
\caption{ Redshift distribution, $N(z)$, of simulated sources
(histograms) compared to input model (solid-lines).  Left-panel shows
$N(z)$ at 850$\mu$m for sources brighter than 0.01, 1.0, 10.0
mJy. Right-panel compares $N(z)$ at 200$\mu$m for sources brighter
than 10 and 100 mJy: The dotted and continuous lines correspond to the
use of Arp220 and M82 as the template to model the galaxy SED.}
\label{selwall}
\end{figure*}

In general, one can recognise the brightest sources in an individual
map at any other wavelength. Consequently it is possible to derive a
robust constraint on the redshift of an individual galaxy from the
relative intensities at different wavelengths ({\it e.g.} Hughes
2000).  Nevertheless, even if we choose the sensitivity limits at each
wavelength to have similar mean redshifts, the redshift distribution
for the $200 \mu$m galaxies is quite different from that of $850 \mu$m
({\it e.g.} see Figure \ref{selwall}).  Therefore care must be taken
in determining the appropriate depths of the complementary surveys
when the goal is to measure millimetre to 
submillimetre colours to estimate the redshift distribution.
These redshift issues will be
quantified in more detail elsewhere (Hughes \& Gazta\~naga 2000).

\section{Source-Count Analysis}

To demonstrate the simulations accurately reproduce the input models
we show in Figure \ref{selwall} the match between the expected number
of galaxies at 850$\mu$m and 200$\mu$m in a given log redshift bin,
$N(z)$, as given by the input selection function, and the measured
counts in the redshift simulations (before including the noise or the
finite resolution to the map).  Figure \ref{selwall} also shows a
comparison of the results for the brighter counts using two different
SEDs models: Arp220 and M82. As shown in Figure \ref{SED}, the
difference in the SED models is larger at $\lambda > 60 \mu$m, so that
for mean redshifts of $z \simeq 2-3$, the choice and/or evolution of
the galaxy SED will only affect the counts at the shortest sub-mm
wavelengths ({\it e.g.} $300-200\mu$m). The right-hand panel of
Figure \ref{selwall} illustrates that at $200\mu$m the effects of
using different SED models on the bright counts ($\sim 100$\,mJy) are
small.  The results at fainter $200\mu$m fluxes or at other
wavelengths are not shown as the differences in the histograms are
indistinguishable for different SEDs.  However the effect on the
counts of the brightest $200 \mu$m sources is also negligible compared
to the sampling effects due to clustering discussed in
\S\ref{sec:clustering}.

The number counts and photometry of the sources in our simulated
angular maps were determined using SEXTRACTOR (within the STARLINK
package GAIA).  In Figure \ref{counts} we compare the predicted and
measured number-counts of the sources extracted from the full
1-deg$^{2}$ simulations at 1100, 850 and 200$\mu$m.  Again there is
excellent agreement between the simulated source-counts and the model
down to the confusion flux limit of the different surveys at their
respective resolutions. This illustrates the obvious point that
provided the survey is of sufficient area and sensitivity then neither
resolution, projection or noise are very important in extracting
counts for objects above the confusion-limit.

\subsection{Resolution and confusion}

\begin{figure}
\centerline{\epsfig{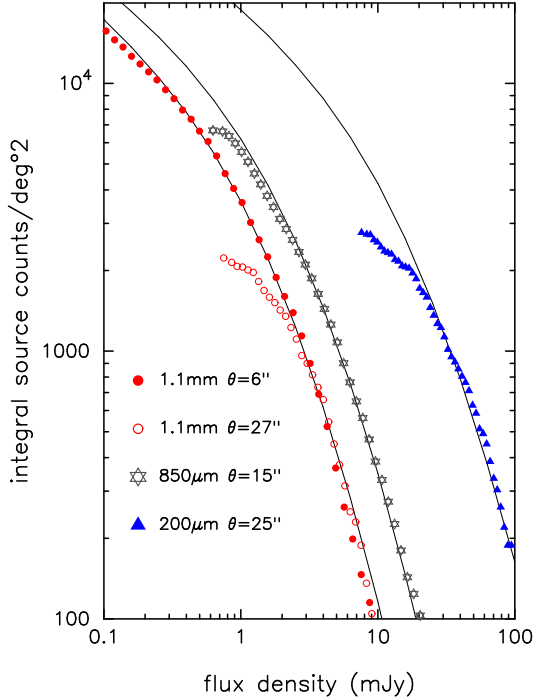}}
\caption{A comparison of the input model (solid-lines) 
and the extracted source-counts 
from the simulated 1-deg$^{2}$ surveys 
at 200$\mu$m, 850$\mu$m and 1.1\,mm. 
Note how the simulated counts are depleted with respect to the input
model at fluxes fainter than the confusion limit, which is set by the
spatial resolution of the survey. The 1.1\,mm counts at different resolutions
(6\arcsec and 27\arcsec) are also shown.}
\label{counts}
\end{figure}

Below the confusion limit  the counts flatten as faint sources
merge to form brighter objects (see  Figure \ref{counts}).
 For example, at 850$\mu$m the counts are affected by
confusion at $S_{850\mu m} \leq  2$\,mJy, whilst at 1.1\,mm with
6\arcsec resolution ({\it e.g.} LMT) 
the counts are still unaffected at $S_{1.1\rm mm}
\sim 0.1$\,mJy.  Even at lower resolution (27\arcsec) the measured
1.1\,mm source-counts recover the input model down to $S_{1.1\rm mm}
\sim 3$\,mJy due to the low density of intrinsically luminous sources
($> 10^{12} L_{\sun}$). Similarly the 200$\mu$m counts recover the
model down to a confusion limit of 18\,mJy at 25\arcsec resolution.
Hence, even at resolutions of $\sim 30\arcsec$ the overlapping of
individual source PSFs is negligible, regardless of whether it's due
to the random line-of-sight projection of galaxies at different redshifts, or
because of a high-amplitude of clustering.  The significance of
this result is that it is therefore possible to combine large-aperture
(50-m LMT, 15-m JCMT) 3000--450$\mu$m wavelength surveys and
relatively small primary aperture ({\it e.g.} $\sim 2.0$\,m BLAST) 
sub-mm and FIR surveys ($500-200\mu$m) with very different
resolutions ($\theta_{FIR}/\theta_{\rm mm} \sim 5$) and still derive
meaningful colours of sub-mm-selected galaxies, and hence 
constrain their redshifts 
(Hughes 2000).

\subsubsection{Extra-galactic Background} 

The 1.1\,mm extra-galactic background with flux
$S > 0.5$ mJy 
due to the discrete sources in our simulations
is $1.31 \times 10^{-10}\rm \, 
W m^{-2} sr^{-1}$, in excellent agreement 
with FIRAS (Far Infrared Absolute Spectrophotometer) results
from COBE
residual measurement of the extra-galactic
far infrared  background ({\it e.g.} Dwek \etal\ 1998
and references therein). This  implies that
surveys conducted at the full spatial resolution of the 
LMT could resolve $\sim 100\%$ 
of the mm background, whilst the deepest SCUBA surveys todate have resolved
only 30-50\% of the sub-mm background.

\subsection{Depth and noise}

\begin{figure}
\centerline{\epsfig{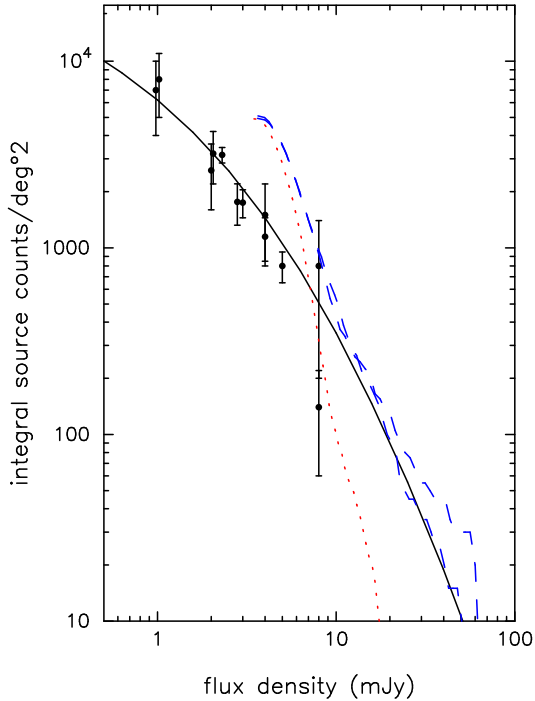}}
\caption{Extracted number-counts at 850$\mu$m from two (3$\sigma \sim
8$\,mJy) 0.2\,deg$^{2}$ simulated surveys (dashed-lines).  The symbols
show the actual observed SCUBA 850$\mu$m number-counts from the
surveys described in \S1 and the solid line correspond to the the
model described in \S2 (which is the input to the simulations).  The
counts from the noise map (1$\sigma = 2.5$\,mJy), which was added to the
raw simulation, are shown as a dotted line. Because these are not deep
surveys, the noise dominates the counts producing spurious simulated
sources at 8-12\,mJy. This has the opposite effect to the confusion
produced by the finite resolution shown in Figure \ref{counts}.}
\label{fig6}
\end{figure}

In the previous subsection we indicated that provided surveys have
sufficient sensitivity that they can reach their confusion limit then
accurate source-counts can be measured from large-area surveys.
However the simulations also illustrate the way in which the counts
are severely affected for shallow wide-area surveys with a flux-limit
well above the confusion limit when the number-counts are steep.  For
example Figure \ref{fig6} shows the simulated counts from a
0.2\,deg$^{2}$ 850$\mu$m SCUBA survey with a 1-$\sigma$ sensitivity of
2.5\,mJy. The extracted counts accurately reproduce the model down to
a flux density of $\sim 12$\,mJy, below which a steep increase in the
counts is observed due entirely to the noisy background.  Hence
individual sources identified in shallow sub-mm surveys at the
3--5$\sigma$ level may be spurious. These additional sources can have
a non-negligible positive contribution to the overall number counts
(compared to the situation in \S3.1 where the counts flatten at the
confusion-limit of the survey).  This is due to the steep decrease in
the galaxy counts which results in an increasing contribution of the
noise counts as the surveys move to brighter flux limits.  As the
noise spectrum is well determined ({\it e.g.}  dashed-line in Figure
\ref{fig6}) a correction can be applied to remove the excess counts
(Hughes \& Gazta\~naga 2000).

\subsection{Clustering and shot-noise}
\label{sec:clustering}

\begin{figure}
~\epsfig{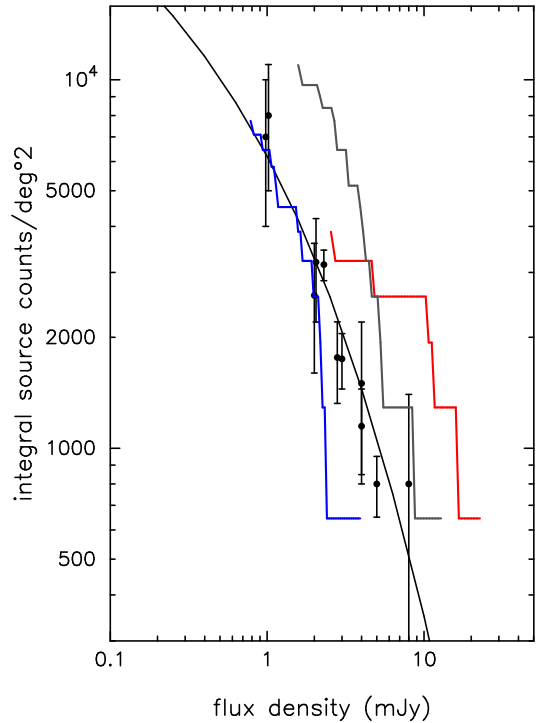}
\caption{Histograms representing the 850$\mu$m source-counts from three
different small area surveys ($\sim 6$\,arcmin$^{2}$) as shown by the
small boxes in Figure \ref{large850}. These are similar in area and depth to
SCUBA 850$\mu$m survey of the HDF (see Figure \ref{hdf}).  
The solid-line and data are the same as in Figure \ref{fig6}.}
\label{fig7}
\end{figure}

\begin{figure}
\centerline{\vspace{8cm}}
\includegraphics{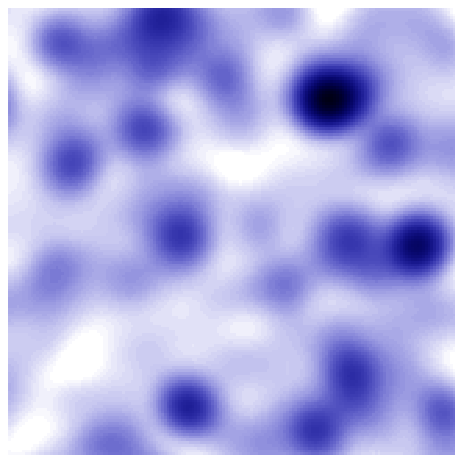}
\includegraphics{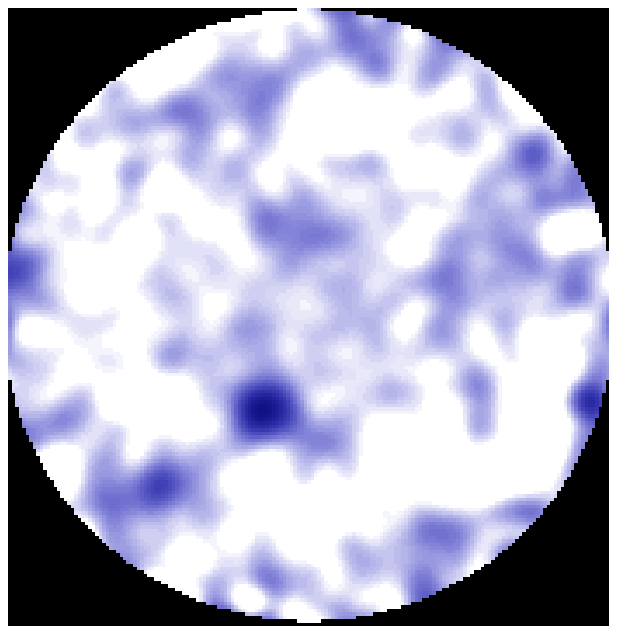}
\includegraphics{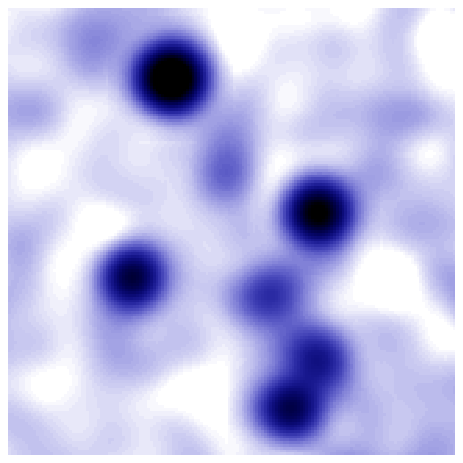}
\includegraphics{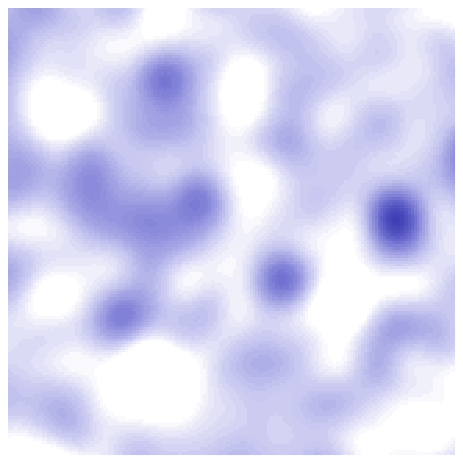}
\caption{Three deep 850$\mu$m surveys extracted from the simulation 
in Figure \ref{large850} with same area and depth as the 
SCUBA 850$\mu$m survey of the HDF (shown  as a circle inside a black 
square for comparison).}
\label{hdf}
\end{figure}

Another important aspect illustrated by the simulations is the effect
induced on the counts by the sampling variance of the large-scale
galaxy clustering.  This is most significant in the deepest surveys
which necessarily provide the smallest survey areas ($<
0.1$\,deg$^{2}$). In Figure \ref{fig7} we compare 3 surveys identical
in area ($\sim 6$\,arcmin$^{2}$) to the SCUBA surveys of the Hubble
Deep Fields (Hughes \etal\ 1998) and similar to those of the Hawaii
Deep Fields (Barger \etal\ 1998) and lensing galaxy clusters (Smail
\etal\ 1997).  We find a factor of 3--10 variation in the extracted
counts from these deep confusion-limited surveys.  The three simulated
deep surveys and the original SCUBA HDF survey are shown in Figure
\ref{hdf}. For the SCUBA surveys we would need an area over $100$
times larger if we want to reduce the variance in the counts to the
few percent level. Note that besides the intrinsic clustering,
shot-noise plays an important role in this variance, at least for the
brighter (less numerous) sources. A detailed analysis of this will be
presented elsewhere (Hughes \& Gazta\~naga 2000).

\section{Conclusions and Future Work}

We have presented simulated surveys which are made as realistic as
possible in order to address some key issues confronting existing and
forthcoming millimetre and submillimetre surveys. These simulations
assume a model for the luminosity and clustering evolution which is
poorly constrained by current data. Improving our understanding of
this evolution provides the motivation for conducting submillimetre
and millimetre surveys. It is therefore important that we quantify the
limitations of these surveys and the significance of the results drawn
from them.

In this paper we discuss the results from survey simulations with a range of
wavelengths (200$\mu$m -- 1.1\,mm), spatial resolutions (6\arcsec --
27\arcsec) and flux densities (0.01--310 mJy).
Several issues have been addressed related to the estimation of 
the measured source-counts in future surveys: 
resolution and confusion; survey sensitivity and noise; and
sampling variance due to clustering and shot-noise. 

The redshift information, determined from the millimetre and submillimetre
colours of individual sources (Hughes 2000), is an essential
ingredient to discriminate between the possible evolutionary models
that describe the star formation history of galaxies.

Clustering in future wide-area submillimetre
surveys might also help to understand some of these issues.
Both the number counts and
the two-point clustering pattern depend strongly on the cosmological model,
but its combination can break some degeneracies. For example
if galaxy formation occurs in the rare peaks of the underlying matter
field then the resulting  galaxy clustering should be stronger
than if galaxy formation occurred in random places.
The measure of the coherence of the clustering pattern, as described by 
the higher-order statistics (which is apparent in our simulated maps),  
can also be used to test how star forming galaxies trace
the underlying mass ({\it e.g.} Gazta\~naga 2000, and references therein). 
This issues will be address in more detail elsewhere.


\end{document}